# Compact Ion Beam System for Fusion Demonstration

Allan X. Chen [a,*], Nai-Wei Liu [c], Alexander Gunn [b], Zhe Su [a], Benjamin F. Sigal [a], Matthew Salazar [b], Nawar Abdalla [b], James Chen [a], Alfred Y. Wong [b], Qiong Wang [a]

Alpha Ring International, Ltd.

[a] 1631 W. 135th St. Gardena, CA 90249, U.S.A.

[b] 5 Harris Ct. Building B-2, Monterey, CA 93940, U.S.A.

[c] 9F., No.11, Ln.35, Jihu Rd. Neihu Dist. Taipei City 114066, Taiwan

**Abstract**

We demonstrate a compact ion beam device capable of accelerating $H^+$ and $D^+$ ions up to 75keV energy, onto a solid target, with sufficient beam current to study fusion reactions. The ion beam system uses a microwave driven plasma source to generate ions that are accelerated to high energy with a direct current (DC) acceleration structure. The plasma source is driven by pulsed microwaves from a solid-state radiofrequency (RF) amplifier, which is impedance matched to the plasma source chamber at the S-band frequency in the range of 2.4-2.5 GHz. The plasma chamber is held at high positive DC potential and is isolated from the impedance matching structure (at ground potential) by a dielectric-filled gap. To facilitate the use of high-energy-particle detectors near the target, the plasma chamber is biased to a high positive voltage, while the target remains grounded. A target loaded with deuterium is used to study D-D fusion and a $B_4C$ or $LaB_6$ target is used to study p-$^{11}$B fusion. Detectors include solid-state charged particle detector and a scintillation fast neutron detector. The complete ion beam system can fit on a laboratory table and is a useful tool for teaching undergraduate and graduate students about the physics of fusion.

**Introduction**

With recent breakthroughs in nuclear fusion technology, such as the announcement of fusion energy break-even by the National Ignition Facility (NIF) at Lawrence Livermore National Laboratory (LLNL) [1], the pace to commercialize fusion has just accelerated. Both research labs and private companies around the world are racing to be the first to demonstrate the commercial viability of fusion energy. At the same time, there needs to be a pool of talent available to sustain the accelerated pace of the R&D needed to make this happen.

Accelerator based ion beam systems have been developed for almost a century [2]. Many laboratories and research institutions around the world house large accelerators that can produce MeV beams or higher of different ions ranging from protons to heavy metallic ions [3][4][5][37].

* Corresponding author at: 1631 W. 135th St. Gardena, CA 90249, U.S.A.
*E-mail address:* allan@alpharing.com (A.X. Chen).

However, the large footprint of such accelerator systems is costly to procure and maintain. There is typically a high demand for use of higher-energy ion beam systems, and so access is limited to students involved in advanced projects and involves scheduling months in advance for a relatively short period of use.  Additionally, the large size and complexity makes it difficult for students to truly interact with the hardware in a hands-on environment.  To that end we have developed a compact ion beam system with the aim of demonstrating various science and engineering concepts that will be relevant for the fusion energy industry of the future.  The core concept of the ion beam system is to accelerate a beam of light ions, either $H^+$ or $D^+$, to bombard a solid target of either D or B to initiate fusion reactions [6][7][35][36][39] that are then detected using various energetic particle detectors.  The compactness of the entire system is ideally suited for deployment in an educational setting where students can have first-hand experience using and interacting with the tool.

**Ion Beam System Design**

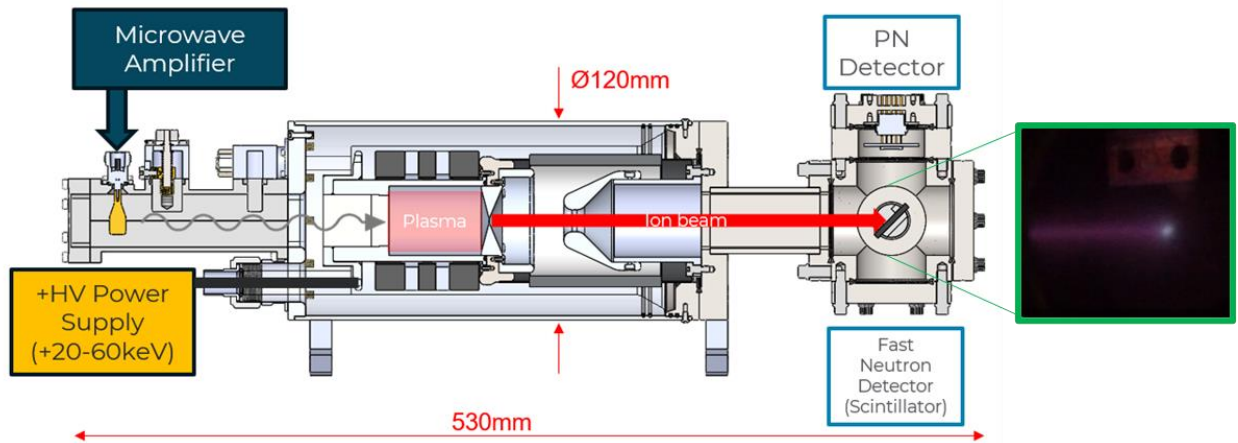

*Figure 1: Schematic and details of the ion beam system interior.  Plasma is generated by ECR microwave excitation.  The ion source itself is floating at potential and ions are extracted into a grounded target chamber to produce fusion particles. Picture on the right shows the ion beam bombarding the target, creating a heated impact zone.*

Fig. 1 shows the basic schematic of the Alpha Ring (ARI) ion beam system (ARI-IBS). Plasma is generated by microwave electron cyclotron resonance (ECR) excitation tuned for the range of 2.4-2.5 GHz.  This is a standard frequency band for which the Federal Communications Commission (FCC) has allocated for unrestricted use for various commercial, scientific, and industrial purposes.  For an ECR plasma, an axial B-field must be present in the excitation region that matches the ECR frequency given by:

$$\omega_{ce} = \frac{qB_z}{m_e} \quad (1)$$

At 2.45GHz, this corresponds to ~900 Gauss. In traditional ECR plasma sources, the B-field is usually produced by an electromagnet or in some cases, a superconducting magnet. These have the advantage that the B-field is very stable and uniform as well as tunability to match the specific excitation frequency. However, the drawback is that they are energy intensive and take up a significant footprint to house the auxiliary equipment (power supplies, cryogenic systems). Therefore, it is difficult to deploy such a system in a compact form factor. The ARI-IBS instead uses permanent magnets to generate the B-field. Modeled after the Peking University ECR ion source [8] and further developed by Adelphi Technology [9], the ARI-IBS uses a set of ring-shaped NdFeB rare earth magnets to generate an axial B-field inside the plasma region, which is roughly 4-cm dia. x 5-cm length. The ring magnets are approximately 10cm dia. with an annulus that fits the ion chamber. The entire size of the ion chamber is no larger than 12cm dia. x 8cm length. The walls of the ion chamber are fabricated from regular 6061-T6 aluminum for good heat dissipation.

The microwave power is coupled into the ion chamber using a ceramic window with low microwave attenuation. This window serves the purpose of microwave transmission as well as vacuum-atmosphere separation. The ceramic comprised of high purity alumina ($Al_2O_3$) is bonded to the aluminum using a special low-outgassing epoxy rated for vacuum operation. On the vacuum side, the ceramic window is directly exposed to hydrogen or deuterium plasma. Over time, the surface of the window could be coated with various metallic materials (mostly Al) from the chamber walls. However, due to the low pressure of operation (~few mTorr), this is a rather slow process, thus giving a lifetime for the ion source of thousands of hours before requiring maintenance or replacement. The entire system is kept under vacuum pressures using a combination of a turbo-molecular pump and a roughing pump. A mass flow controller (MFC) is used to inject a small amount of either $D_2$ or $H_2$ into the system depending on whether the user wants to observe the D-D or p-B11 reaction.

On the atmospheric side of the window, a dielectric material is placed in direct contact with the ion source body for high voltage standoff from the grounded waveguide structure. This is typically a plastic with very low microwave attenuation (e.g. polyethylene) to permit good microwave power coupling into the ion source. The grounded waveguide structure consists of a two-stub impedance match tuning element and a coaxial-to-waveguide adapter to facilitate the optimal power transfer between the microwave amplifier and the plasma load. The two-stub impedance match is typically tuned during initial operation to find the minimal reflected power (corresponding to maximum forward power into the plasma). For initial coarse tuning, we first attach a network analyzer to the coaxial-to-waveguide adapter to find the minimum in the $S_{11}$ scattering parameter. Typically, the tuned frequency will be 2.45 GHz as the amplifier output is optimized for this frequency.

After the plasma is generated, the $H^+/D^+$ ions are accelerated through a single gap extraction electrode down the high voltage column towards the target chamber. The ion beam drift distance is ~18cm to minimize particle collision with the background gas, which is kept in the mTorr range. For D$^+$ beam operation, the target typically comprises an aluminum disk with a deuterated-diamond-like-coating (DDLC) to maximize deuterium density on the surface. For H$^+$ beam operation, the target usually consists of LaB$_6$ or B$_4$C. Both have relatively high stoichiometric ratio of boron in solid density. Using a viewport mounted in the target chamber, one can observe the ion beam impinging on a metal target at 30 keV (Fig. 1). The white spot at the end of the beam marks the beam target interaction zone.

Back-streaming electrons pose a major issue in our compact ion beam for two reasons: (1) They increase the current draw of the power supply by a factor of 2 times the beam current. (2) The electrons emitted from the target tend to be accelerated back towards the plasma region, potentially causing damage and reliability issues during long term operation. We mitigate this issue by adding a small piece of permanent magnet in the ion beam drift region to bend the electron trajectories and cause them to be dumped into the chamber wall before they are accelerated to high energies through the single gap extraction region. The magnetic field strength in the vacuum region is ~1 kilogauss and the field direction are perpendicular to the ion beam propagation direction.

**Ion Beam Experimental Results**

The compact ion beam has been successfully operated at 30 keV with D$_2$ pressure at 8 mTorr. The microwave input power is at +3 dBm using the TPI-1001-B RF synthesizer with voltage-controlled oscillator (VCO) tuned to 2.45 GHz [10]. The RF amplifier is a ZHL-2425-250+ model produced by Mini-Circuits which has nominal gain of 42 dB in the range of 2.4 – 2.5 GHz [11]. Thus, the nominal microwave power input into the ion source with ideal impedance matching is only ~31.6 W. Impedance matching is performed using the network analyzer as discussed in the previous section. A smith chart and S$_{11}$ plot can be obtained as a function of frequency when the impedance matching has been tuned for 2.45 GHz operation. Due to the resolution of the network analyzer, the value of S$_{11}$ at the minimum can fluctuate by about one decade. However, the minimum should be < 30 dB relative to the baseline to be considered tuned. Fine tuning can be performed with a plasma load and/or ion beam to maximize the beam current. It is worth noting that since the accelerator column is filled with a dielectric fluid TMC-3283E [12] during operation for cooling and electrical isolation purposes, the tuning will change because of the difference in dielectric constant similar to that of a loaded waveguide. Therefore, final tuning should be done with the dielectric fluid running to have the best impedance matching condition.

Particles of fusion products have been obtained with the ion beam operating at 5% duty cycle and 30 keV acceleration with tuned impedance at 2.45 GHz. The ion beam pulsing is accomplished by pulsing the microwave power. The typical pulse frequency can range from ~1 Hz to 1 kHz, thus providing a beam pulse length between 50 µs to 50 ms. The nominal current observed on the acceleration voltage power supply is ~0.048 mA. We used an aluminum target with a thin 3-5μm layer of deuterated-diamond-like-carbon (DDLC) coating to enhance the surface density of deuterium. As discovered using a cloud-chamber setup like the one shown in Fig. 2, this allows a much higher fusion particle yield of approximately 4X compared to the bare aluminum target. The cloud chamber setup consists of a thin aluminum foil membrane (25 μm) located at one of the target chamber ports which is 90º normal to the beam propagation axis. This thin material allows the 3.0 MeV protons to pass through while shielding out most of the 0.82 MeV helion and 1.01 MeV triton nuclei due to the lower stopping power of protons in aluminum. On the DDLC coated side, the neutron yield is ~200 n/s while on the uncoated side, the yield is ~50 n/s as measured by a helion thermal neutron detector. The cloud chamber is a standard method of visualizing the fusion particle tracks in atmosphere [13][14].

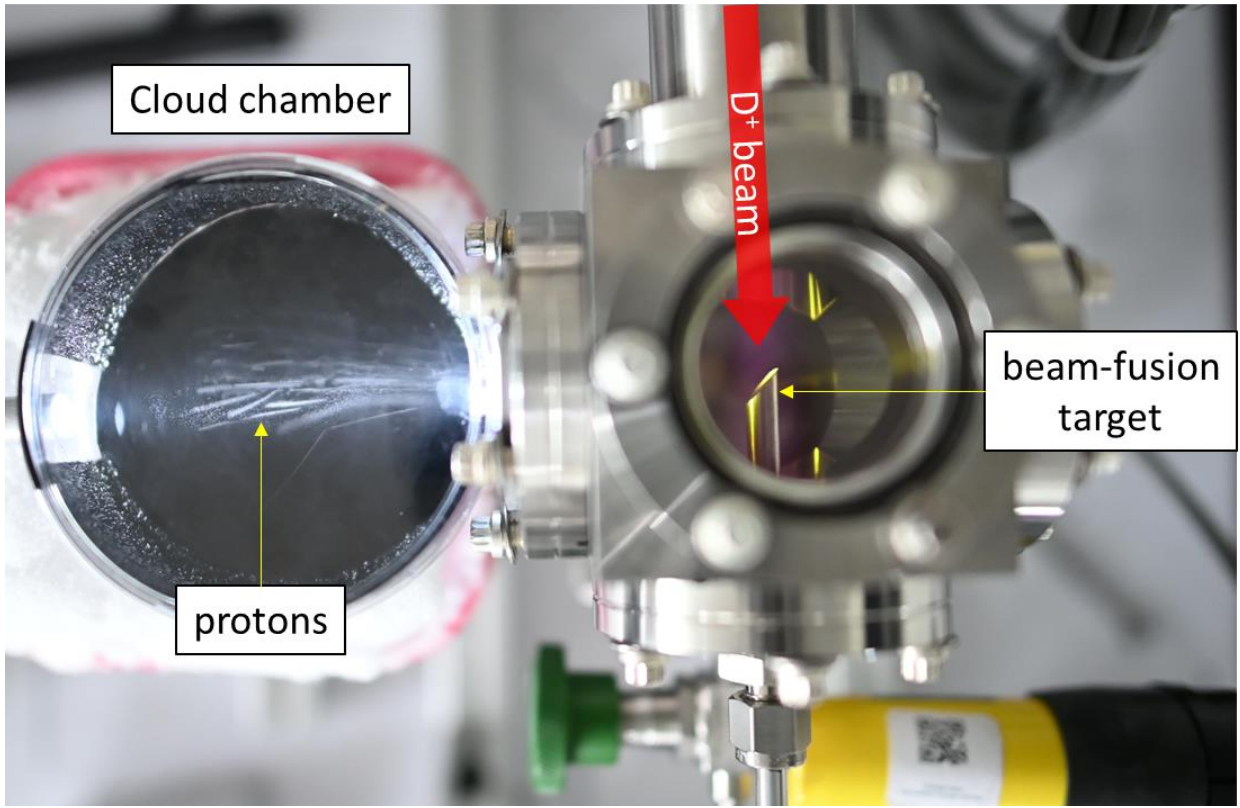

*Figure 2: A photograph of the ion beam striking an aluminum target. Tracks of protons from D-D fusion can be observed in a supersaturated ethanol cloud chamber after passing through a 25 μm Al foil membrane. The ion beam impinges on the target surface, which is oriented 45° to the beam and the Al foil membrane, allowing protons to be observed.*

In addition to visualizing the fusion reactions using the cloud chamber, we have also used CR-39 coupons for nuclear track detection. This is a standard method [15][16][22] to observe charged particles from both D-D and p-B11 fusion. The use of CR-39 can be more immune to RF, light, and other electronic interference than solid-state energetic-particle detectors (e.g. PIN or PIN photodiodes). Fig. 3a shows the CR-39 detection setup for measuring D-D fusion particles. We used several different thicknesses (0, 4, 8, 12, and 16 μm) of Al foil to range out the three different charged particles. The CR-39 coupon obtained from Track Analysis Systems

Ltd.[34] (TASL) with dimensions of 20 mm x 25 mm x 1.6 mm is mounted on a standard CF-2.75 stainless steel blank flange inside the target chamber.

Once the glass window flange is mounted into the target chamber, the system is pumped down to less than 1E-5 Torr before ion beam operation. The pump down process typically takes about 1-2 hours depending on the level of atmospheric exposure of the vacuum components. Ion beam operating conditions follow the same parameters as described in the previous section. The total irradiation time was 10 minutes for D-D in order to collect enough tracks on the CR-39 for analysis. After irradiation, the CR-39 coupon was etched using 6 M NaOH at 90°C for 3 hours. Preliminary results of the CR-39 tracks are shown in Fig. 3(b - d). For the D-D reaction, the three groups of particles are clearly seen in the section without foil covering. The 3.02 MeV protons leave the smallest tracks amongst the three particles due to their higher energy and lower mass. The tritons leave the largest tracks due to their slightly higher energy compared to the helions (1.01 MeV vs. 0.82 MeV). Additionally, the triton track contrast is also much sharper by comparison. With different thicknesses of aluminum foil, the helion and tritons can be ranged out respectively, leaving only the proton tracks visible at the largest thickness of 16μm.

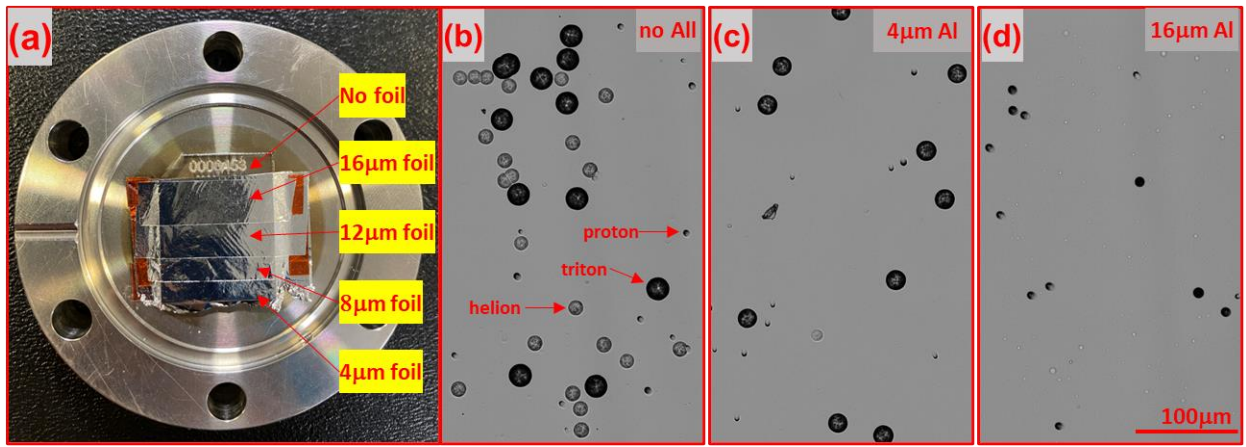

Figure 3: (a) Nuclear track images of D-D fusion particles on CR-39 (coupon#453) with different thicknesses of Al foil covering. Images were taken under a microscope with 40x magnification. The ion beam parameters were 30 keV/0.03 mA, 5% duty cycle, 1 kHz at 3.6 mTorr pressure for 10 minutes of irradiation. (b) With no Al covering, all three particles can be identified. (c) With 4 μm Al foil, the helion particles are stopped and does not form tracks on the coupon. (d) With 16 μm Al foil, both helion and triton are stopped, leaving only the protons visible.

A similar ion beam irradiation of the CR-39 was performed for p-B11 fusion. Compared to D-D fusion, p-B11 has much lower yields at sub-650 keV incident beam energies. We therefore chose to operate the ion beam at 75 keV for 20 minutes to collect enough tracks for analysis. The right-hand side of the CR-39 coupon (7a) is covered with a 26 μm Al foil, which is thick enough to stop all alphas from p-B11 while the left-hand side was left uncovered for unattenuated alpha bombardment. As a sanity check, we used a known alpha source (Am-241 with no surface metallization) to perform a calibration exposure on another coupon. Both coupons were subject to the same vacuum conditions in order to preserve their CR-39 track damage characteristics. The calibration coupon was irradiated with a 1 μCi Am-241 for 4 minutes immediately after p-B11 alpha irradiation. Preliminary results of the nuclear track images can be seen in Fig. 4. The uncovered left section (7b) shows an even distribution of track sizes, which is consistent with the wide-energy spectrum of p-B11 alpha particles. Additionally, we also see a range of track contrasts, which is also consistent with the alpha energy comprising a distribution rather than mono-energetic peaks.

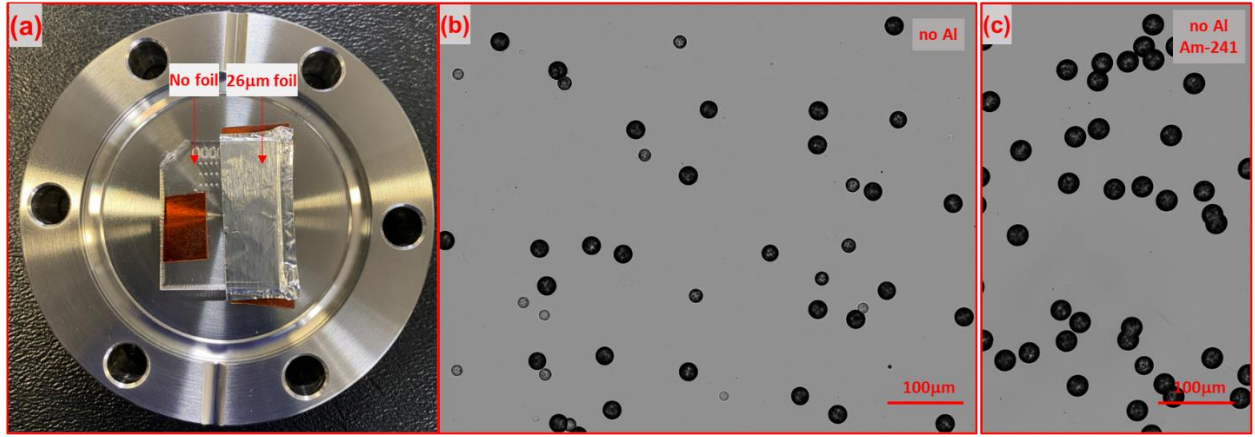

*Figure 4: (a) Nuclear track images of p-B11 fusion alpha particles on CR-39 (coupon#711). Am-241 calibration sample (c) was done as a sanity check. Images were taken under a microscope with 40x magnification. (a) A range of different sized particle tracks are visible, confirming the wide-energy spectrum of the p-B11 alphas. (b) Track density under the 26 um foil is consistent with background. The ion beam operating parameters were 75 keV/0.05 mA at 3.2 mTorr for 20 min. of irradiation. Ion beam duty cycle was set at 5% and pulse frequency set to 1 kHz to limit the average current and thermal load on the target.*

In addition to using CR-39, we also utilized solid-state electronics detectors to capture the charged particles from the D-D and p-B11 reaction. The solid-state detector consists of a large area Hamamatsu Si-PIN photodiode (S14605) which has high quantum efficiency and high energy resolution for radiation detection [17]. In order to block out spurious light from the ion beam, we used a 0.8 μm Al foil in front of the detector to block out any stray light while minimizing attenuation of the charged particle signals [18]. The depletion layer of 500 μm is thick enough to stop all of the charged particles from D-D and p-B11 [19]. The detector operates as a charged coupled device responding to the fusion particles that are impinging on the PIN diode surface. We built a custom JFET-based two-stage pre-amplifier following the conceptual design from Cremat's CR-110-R2.2 charge-sensitive preamplifier [20]. Our choice for the op-amp is the ADA4625-2 due to its low voltage noise characteristics [21]. The entire circuit fits inside a 1" diameter PCB and can be mounted on a CF-2.75 vacuum flange. This allows the preamp electronic circuitry to be connected directly to the PIN detector inside vacuum for minimal stray capacitance and EMI isolation. Fig. 5 shows the basic schematic and design of the solid-state detector. The positive and negative rails of the op-amp are powered by 1S LiPo batteries while the bias voltage is supplied by either one or two A23 alkaline batteries. These batteries are placed outside of the vacuum in a small, shielded aluminum box that can be connected to the PIN detection circuitry via a DB-9 vacuum feedthrough.

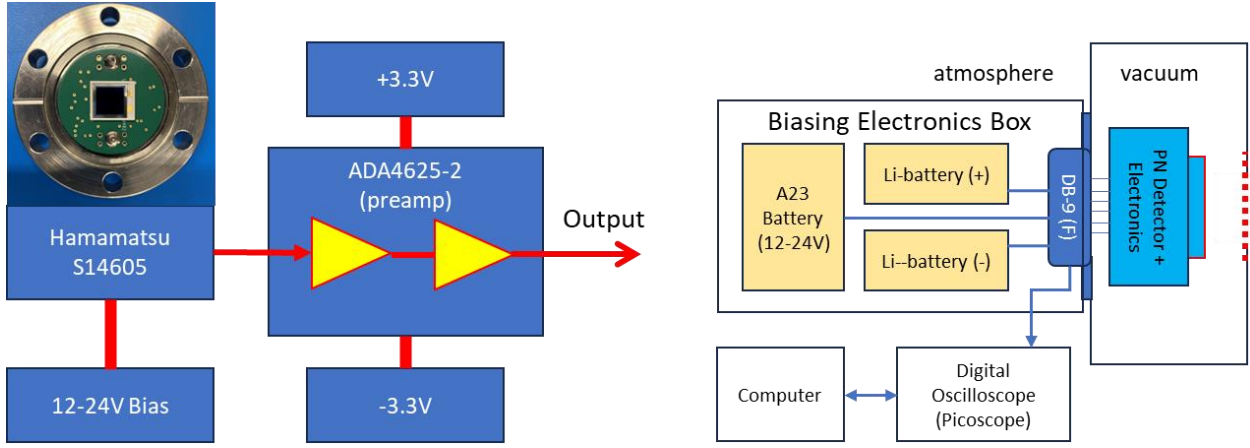

*Figure 5: (left) Electronic schematic of the PIN detector for observing charged particles from D-D and p-B11 fusion reactions. The PIN detector signal is read directly after a 2-stage low voltage noise preamp using the ADA4625-2 op-amp. The signals are then digitally processed to apply various pulse shaping techniques to improve the signal to noise ratio. (right) The preamp electronics are mounted on the atmospheric side of the CF-2.75 flange while the PIN detection unit is mounted on the vacuum side.*

The preamp output from the PIN detector allows us to get very good timing information regarding the charged particle pulses. In order to get a good signal to noise ratio for energy analysis, we applied digital pulse shaping in discrete time to transform the preamp output pulses into semi-gaussian pulses [38]. The pulse shaping algorithm consists of a first order high-pass (CR) followed by four low-pass (RC$^4$) filters. Implementation of the filters was done using MATLAB and the discrete form of the equation is given below.

$$y_2(t) = \alpha_{CR} \cdot y_2(t-1) + \alpha_{CR} \cdot [y_1(t) - y_1(t-1)] \qquad (2)$$

$$y_m(1) = \alpha_{RC} \cdot y_{m-1}(1) \qquad (3)$$

$$y_m(t) = \alpha_{RC} \cdot y_{m-1}(t) + (1 - \alpha_{RC}) \cdot y_m(t-1) \quad ; \quad m = 3 \text{ to } 6 \qquad (4)$$

$$\alpha_{CR} = \frac{\frac{R_H C_H}{\Delta T}}{\frac{R_H C_H}{\Delta T} + 1} \quad , \quad \alpha_{RC} = \frac{1}{\frac{R_L C_L}{\Delta T} + 1} \qquad (5)$$

Here, $y_1(t)$ is the discrete-time dependent vector of preamp signals, $y_2(t)$ is the signal after CR filtering, $y_3(t)$ to $y_6(t)$ are the signals after the 1st to 4th RC filtering, $\alpha_{CR}$ and $\alpha_{RC}$ are the coefficients which relate to the high-pass and low-pass values RC values ($R_H$, $C_H$, $R_L$, $C_L$) and the time step, $\Delta T$. We chose $\alpha_{CR} = 0.85$ and $\alpha_{RC} = 0.15$ with $\Delta T = 64$ ns. The resulting semi-gaussian pulse shaped output has a pulse width of ~3-4 μs (Fig. 6). The peak height of the shaped pulse is directly proportional to the particle energy. By binning the peak heights, one can obtain an energy spectrum of the D-D or p-B11 reaction.

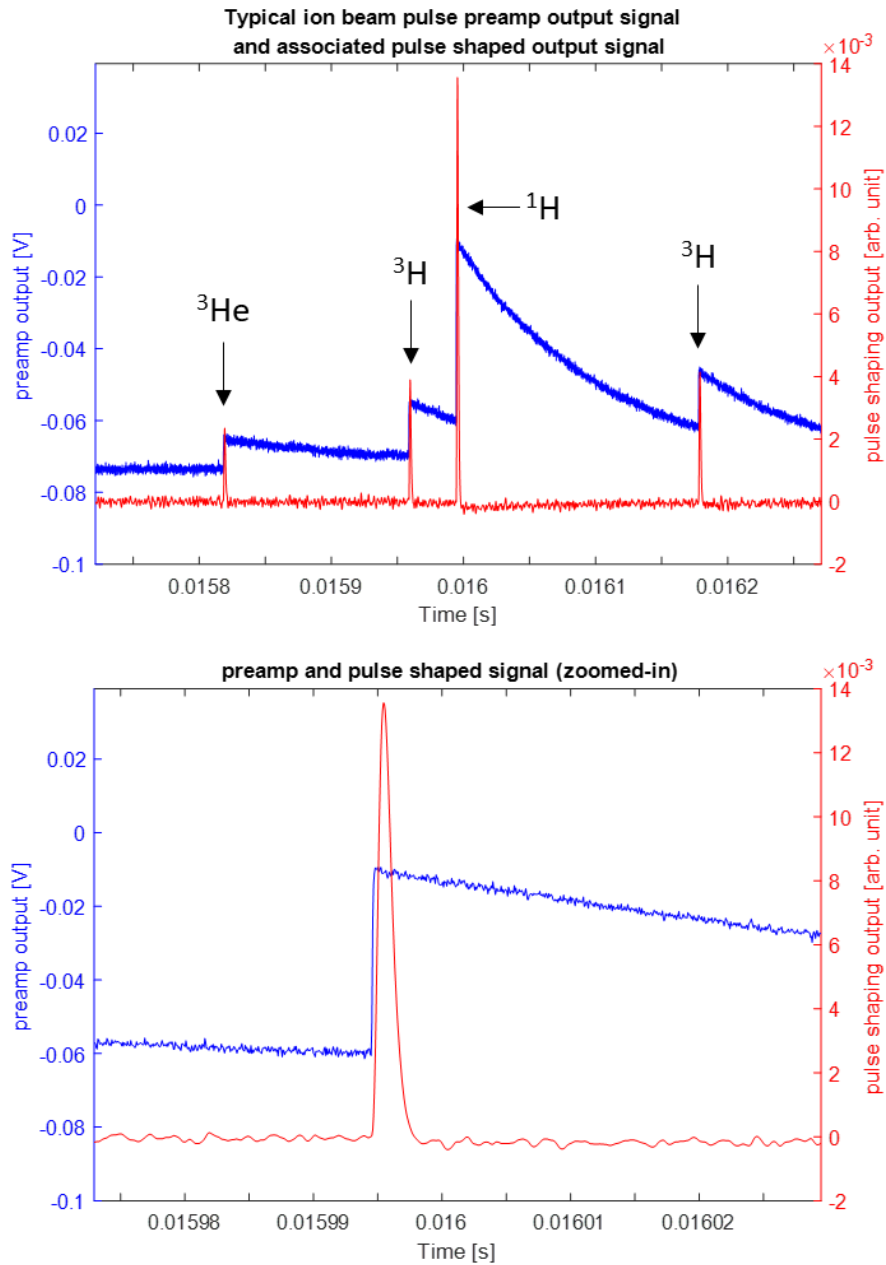

*Figure 6: (Top) Typical pulses from the D-D reaction showing the preamp and pulse shaped signal from different particles. (Bottom) Zoomed-in view of the proton pulse showing the pulse width of the pulse shaped signal.*

The D-D spectrum was obtained at 40 keV with average beam current ~0.033 mA integrated for approximately 4.4 minutes (Fig. 7). The target used is the DDLC coated material discussed in the previous section. The energy scale was calibrated against the proton peak at 3.02 MeV, which has minimal attenuation when passing through the 0.8 μm thick Al foil. The SRIM calculation for stopping range of 3.02 MeV protons in Al is ~80 μm. In the case of alpha particles, the stopping range is much shorter due to having two protons (e.g. 5.5 MeV alphas would stop in 24 μm of Al). The alphas in p-B11, which are mostly in the range of 3-5 MeV, would be attenuated by ~130 - 180 keV in 0.8 μm of Al. To study p-B11 fusion, a spectrum (Fig. 8) was obtained by operating at higher beam voltage (75 keV) and slightly higher average beam current (0.048 mA), as well as longer total irradiation time of 22.2 minutes. This is necessary due to the lower cross-section of the p-B11 reaction [22]. The target is a $B_4C$ material, which can be readily obtained as a sputtering target [23]. The energy scale is the same as that for the D-D spectrum.

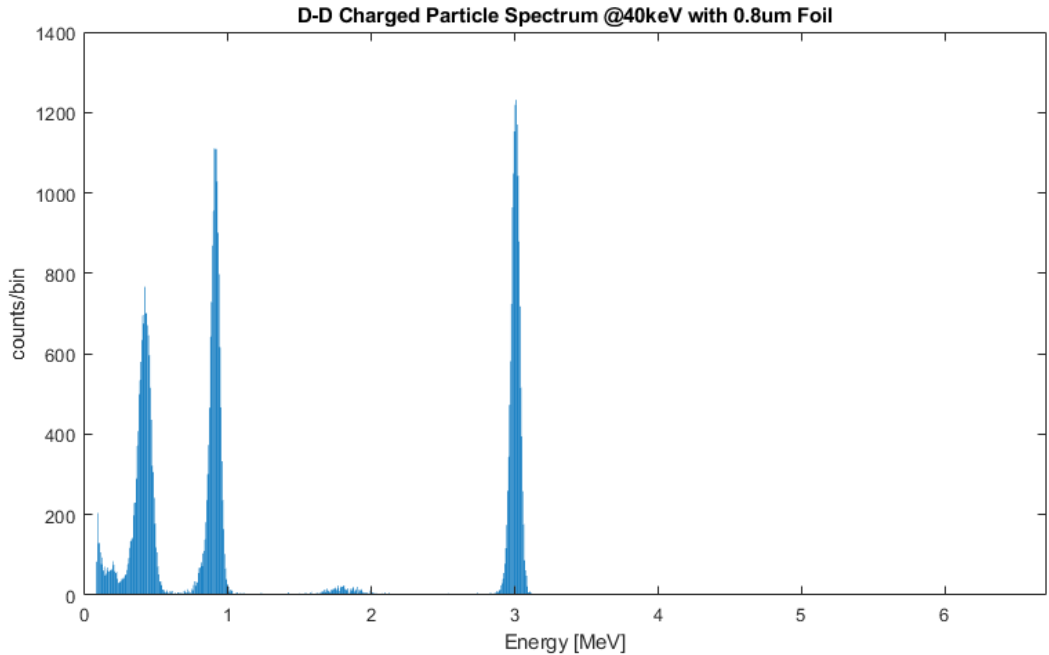

*Figure 7: Charged particle spectrum of the D-D reaction measured by the PIN detector for ion beam operating conditions 40 keV/0.033 mA at 3.3 mTorr pressure. A 0.8 μm foil was placed in front of the PIN detector to block off light from the ion beam and target heating. The three peaks from proton, triton, and helion are clearly identified.*

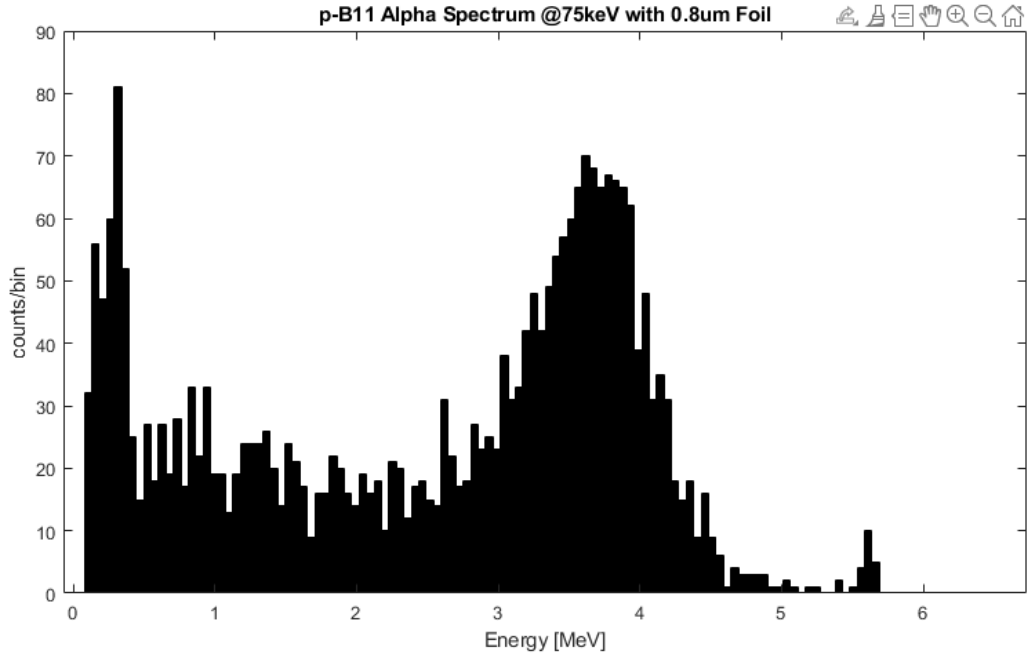

*Figure 8: Charged particle spectrum of the p-B11 reaction measured by the PIN detector for ion beam operating conditions 75 keV/0.048 mA at 3.3 mTorr pressure. A 0.8 μm foil was placed in front of the PIN detector to block off light from the ion beam and target heating. The alpha particle energy spectrum matches well with that obtained from an accelerator-driven p-B11 reaction [30].*

Using the PIN detector and operating at different beam energies, one can obtain a good estimate of the fusion yield as a function of the beam energy (Fig. 9 and 10). Here the fusion yield is defined as the number of particles hitting the detector per second per mA of beam current. We scanned the ranges of 10 to 40 keV for D-D and 45 to 75 keV for p-B11 fusion. The energy ranges were chosen to obtain enough counts for statistical significance on the low end while minimizing pileup effects on the high end. For p-B11 fusion, since the cross-section is orders of magnitude lower than D-D, we need to operate at much higher voltage to obtain statistically significant number of counts. We note here that the beam voltage uncertainty is approximately 1 kV based on the discrepancy between the power supply setpoint and reported value. The error on detector count rate was treated based on a sample size of one for each beam parameter [40]. For both runs, the pressure was kept constant at 3.3 mTorr and the ion beam parameters were also kept constant with the repetition rate of 3 Hz and duty cycle of 5%. The current varies slightly over the range of operating voltage due to higher current being extracted as the electric field increases in the extraction gap.

The ion beam system was operating very stably throughout the entire run at different voltages. To generate the p-B11 fusion yield curve, we had to operate the ion beam continuously for more than 2 hours to generate the data. The p-B11 yield curve shows an exponential dependence while the D-D charged particle yield curve shows a better fit to a power law dependence rather than an exponential dependence. Based on the fusion cross-section, we would expect an exponential dependence with energy. However, for the case of D-D, there is a loading factor which is largely unknown and somewhat difficult to characterize. Given the fact that we only operated for ~4.4 minutes at each voltage, this may not be enough time for the target to reach an equilibrium with the D-loading, which may be the cause for the non-exponential behavior. Both the high voltage and plasma source were stable as observed by the ion beam current.

| E [kV] | error | Detector Tally Rate [Counts/sec] | Beam current [mA] | Current Normalized Detector Tally Rate [Counts/sec/mA] | error | Detector Tally [Counts] | error |
|---|---|---|---|---|---|---|---|
| 10.2 | 1.0 | 0.34 | 0.0277 | 12.2 | 1.3 | 90 | 9.5 |
| 15.3 | 1.0 | 1.40 | 0.0293 | 47.6 | 2.5 | 372 | 19.3 |
| 20.4 | 1.0 | 4.64 | 0.0301 | 154.2 | 4.4 | 1238 | 35.2 |
| 25.5 | 1.0 | 15.83 | 0.0293 | 540.1 | 8.3 | 4220 | 65.0 |
| 30.6 | 1.0 | 42.21 | 0.0316 | 1335.6 | 12.6 | 11255 | 106.1 |
| 35.7 | 1.0 | 89.42 | 0.0324 | 2760.0 | 17.9 | 23846 | 154.4 |
| 40.8 | 1.0 | 164.85 | 0.0331 | 4980.5 | 23.8 | 43961 | 209.7 |

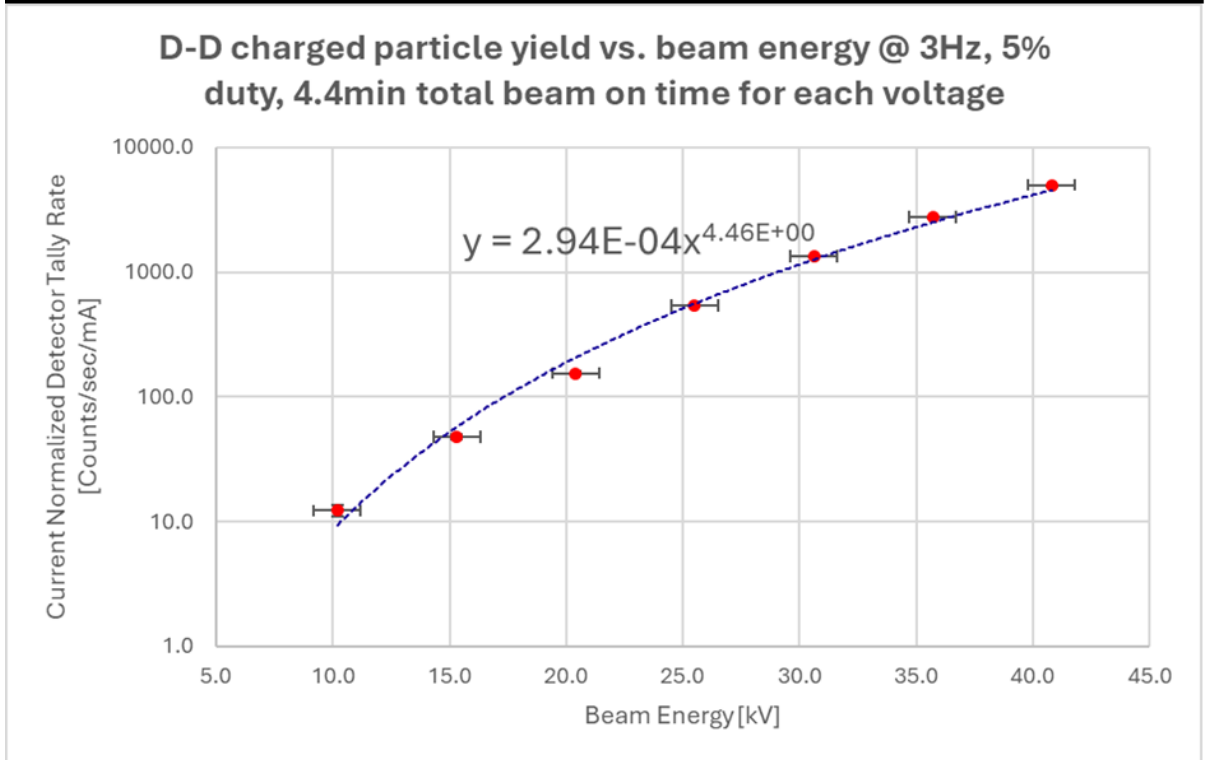

Figure 9: D-D charged particle counts on the PIN detector vs. beam energy. The yield follows a power law dependence on the beam energy.

| E [kV] | error | Detector Tally Rate [Counts/sec] | Beam current [mA] | Current Normalized Detector Tally Rate [Counts/sec/mA] | error | Detector Tally [Counts] | error |
|---|---|---|---|---|---|---|---|
| 46.0 | 1.0 | 0.013 | 0.0393 | 0.3 | 0.1 | 11 | 3.3 |
| 51.1 | 1.0 | 0.038 | 0.0424 | 0.9 | 0.2 | 32 | 5.7 |
| 56.2 | 1.0 | 0.060 | 0.0424 | 1.4 | 0.2 | 50 | 7.1 |
| 61.3 | 1.0 | 0.143 | 0.0455 | 3.1 | 0.3 | 119 | 10.9 |
| 66.5 | 1.0 | 0.392 | 0.046 | 8.5 | 0.5 | 327 | 18.1 |
| 71.5 | 1.0 | 0.920 | 0.047 | 19.6 | 0.7 | 767 | 27.7 |
| 76.6 | 1.0 | 1.980 | 0.0478 | 41.4 | 1.0 | 1650 | 40.6 |

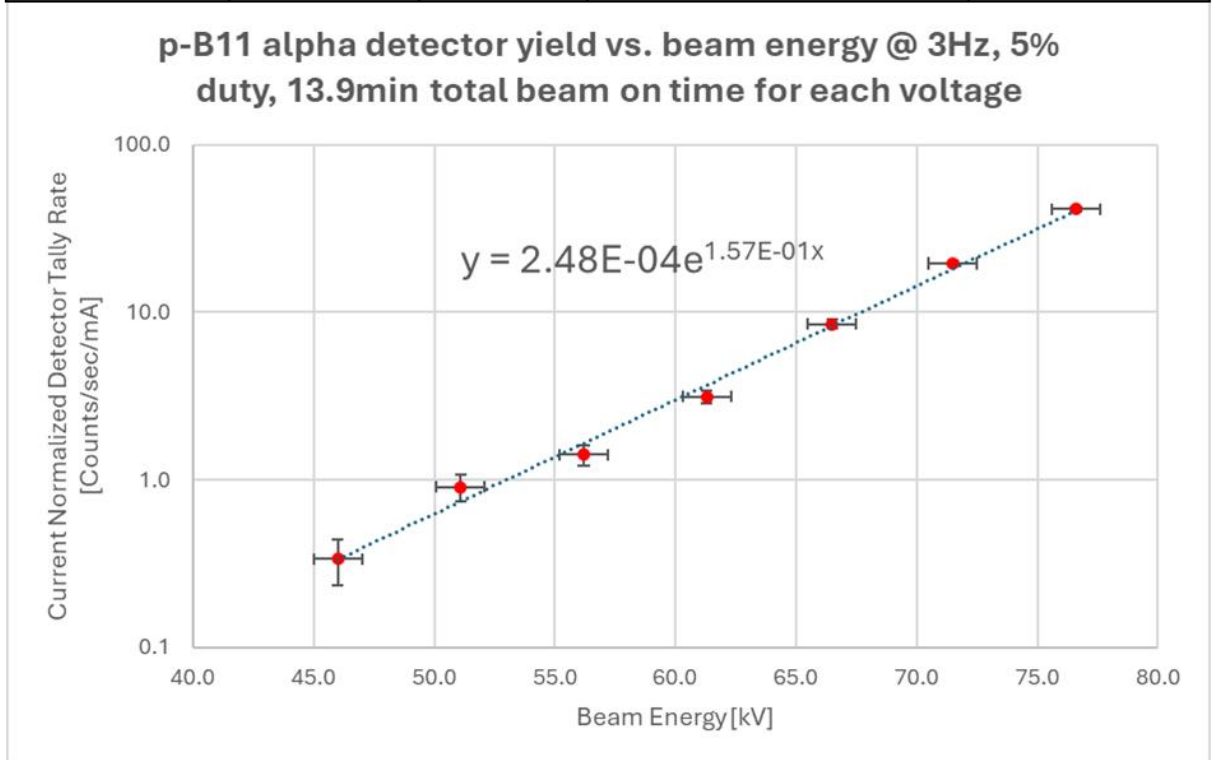

Figure 10: p-B11 alpha particle counts on the PIN detector vs. beam energy. The yield follows an exponential dependence on the beam energy.

For proof of concept, we also characterized the high voltage power supply (HVPS) current with and without the electron suppression magnets to observe the back-streaming electron contribution. Operating with hydrogen at 50 kV, the HVPS current with and without the suppression magnets were 0.0424 mA and 0.1233 mA respectively. Thus, the electron contribution to the current was ~1.9 times of the actual ion beam current.

In certain applications, such as associated particle imaging (API) [33], there's an interest in knowing the beam spot size on the target. A small spot size is preferable as it is closest to the point source. To get a good estimate of the target spot size, we used a new target for the CR-39 (coupon#711) nuclear track detection run. For this, the ion beam was operated only at 75 keV for 20 minutes at 5% duty cycle and 1 kHz with an average current of 0.05 mA (peak current ~1 mA). After the run, the target chamber was vented to atmosphere to inspect the target and the spot size was deduced to be 0.23 cm.

For D-D fusion, the neutron particle is also an important by-product of the reaction. We utilize a plastic scintillator material coupled to a photo-multiplier tube (PMT) to detect the neutrons from the D-D reaction. The plastic scintillator (EJ-276 from Eljen Technology [24]) emits a burst of light whenever an energetic particle, such as neutron or gamma-ray, scatters off the protons in the plastic. This optical pulse is captured and quickly converted into an electronic pulse inside the PMT (Hamamatsu R6231 [25]). The PMT operates in capacitively coupled mode with typical bias voltage of +1000V. A schematic and design of the neutron detector unit is shown in Fig. 11. A cylindrical mu-metal magnetic shield [26] is placed around the PMT to minimize fluctuation in the PMT gain due to stray magnetic fields.

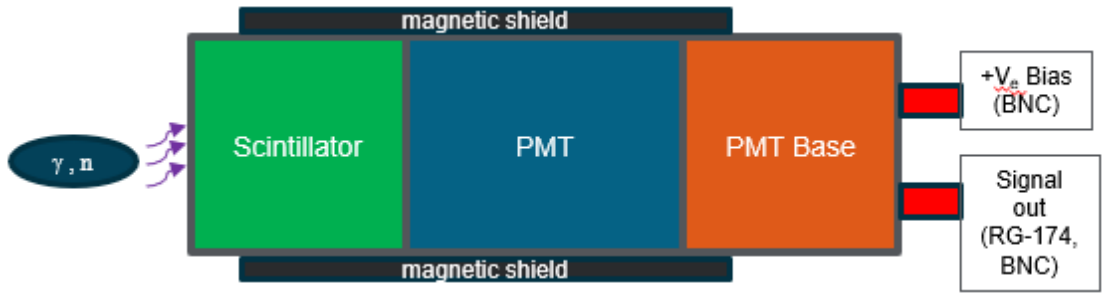

*Figure 11: Schematic of the fast neutron detector which comprises of an EJ-276 scintillator and PMT assembly. The PMT is positively biased to operate in capacitively coupled mode for optical pulse detection. The signal output is connected to a fast digital oscilloscope and terminated with a 500 Ω resistor. The resulting pulse rise and fall times are shown in the next figure.*

For organic scintillators, the optical pulse consists of prompt and delayed fluorescence components which have decay times on the order of 1E-9 and 1E-7 seconds respectively [27]. Neutrons that scatter off the protons in the plastic scintillator will likely produce delayed fluorescence while gamma-rays that scatter tend to produce prompt fluorescence. The difference in optical light decay times between neutron and gamma-ray scatter allows us to separate the signals by pulse-shape discrimination (PSD). Fig. 12a illustrates the difference between a neutron pulse (red) and gamma pulse (blue) after pulse height normalization. Note that the neutron pulse has a slightly longer decay constant compared to the gamma pulse. A typical PSD plot is shown in Fig. 12b, where the x-axis displays the total integral of the signal, and the y-axis displays the partial integral of the signal. Due to the difference in decay time between the neutron and gamma pulse, one can see a separation between the two groups.

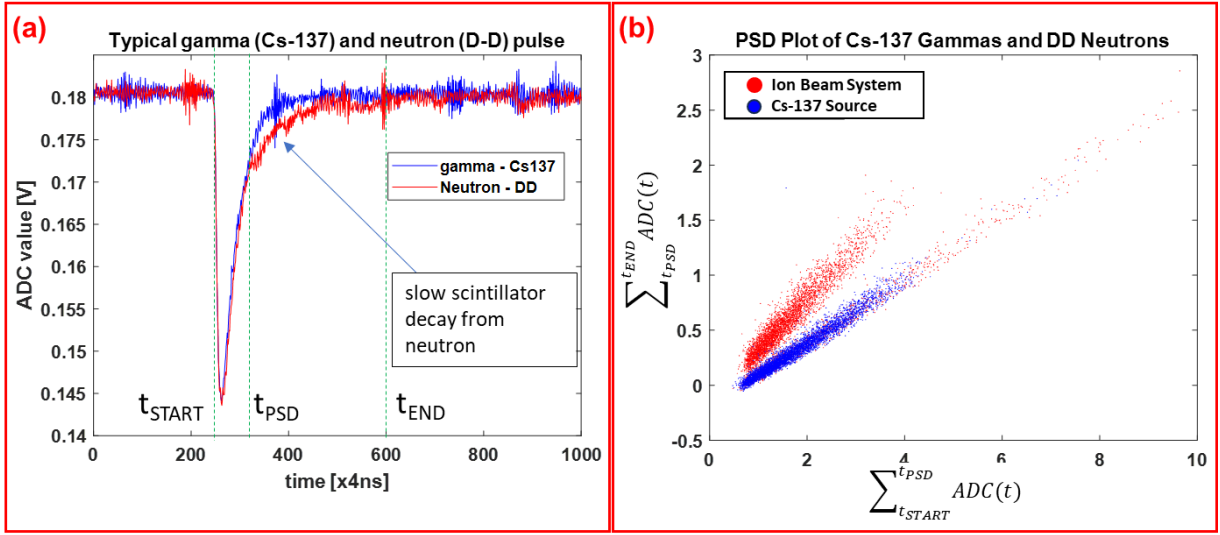

*Figure 12: (a) Typical photomultiplier signal of a Cs-137 gamma [blue] and D-D neutron [red]. We chose two signals with approximately the same peak height to emphasize the slow scintillator decay component present with the neutron signal but not the gamma signal. Here, we chose [tSTART, tPSD, tEND] to be [1000 ns, 1260 ns, 2400 ns] rather arbitrarily to coincide with the observed time window of the slow decay component. (b) Aggregate plot of the Cs-137 gamma and D-D neutron signals based on the chosen PSD parameters, separated groups are clearly distinguishable.*

    Based on the D-D reaction, we also know that the helion and neutron particles are coincident and move in opposite directions due to conservation of momentum. With this knowledge, we can place the PIN and fast neutron detectors at 180 degrees to detect the helion and neutron simultaneously (Fig. 13). We used a faster sampling rate of 250 MSps in order to time resolve the neutron detector signal with good time resolution. This is necessary when performing the PSD algorithms. Notice that during this period, there were two particles detected by the PIN detector, a proton and a helion. Only the helion signal shows coincidence with the neutron detector signal. Using the coincidence technique, one can obtain a set of signals from the fast neutron detector which are guaranteed to be neutron signals. This set of signals can be used to train the PSD algorithm with good confidence.

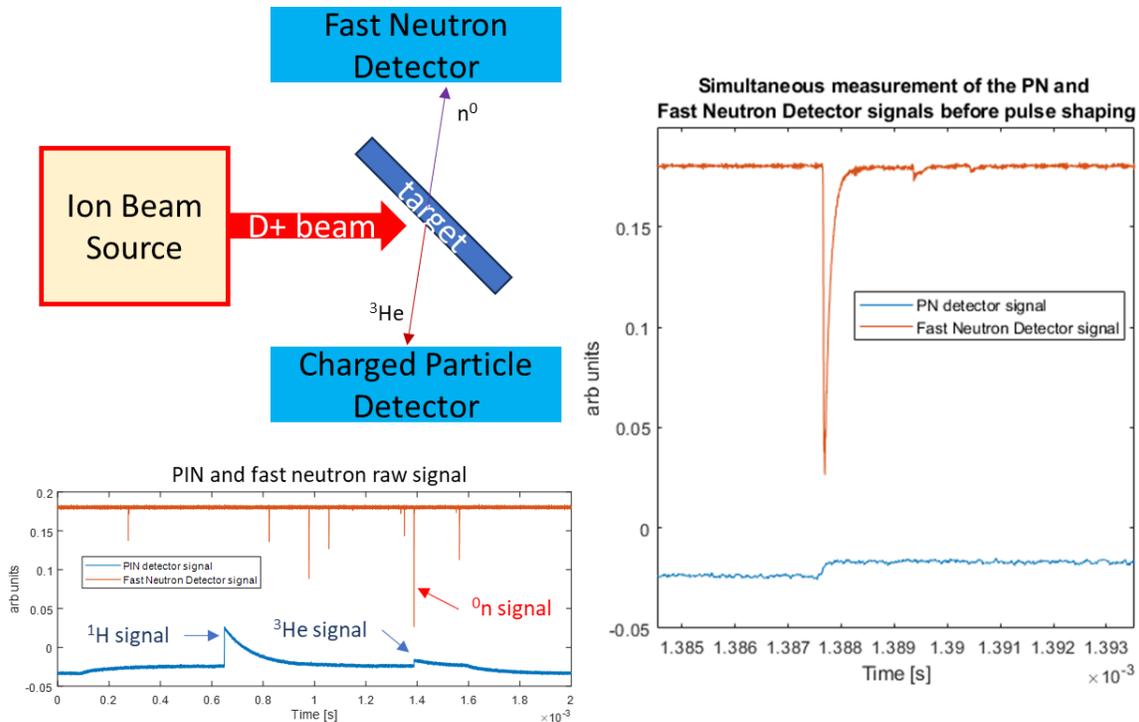

*Figure 13: (Top Left) Configuration for the PIN and fast neutron detector coincidence measurement of the helion and neutron. (Bottom Left) PIN and Fast neutron detector signals. (Right) zoomed-in view of the neutron signal showing that it is time correlated with the helion signal to less than 200 ns. The time correlation may be slightly shifted due to cable length delays.*

**Radiation Safety and Activated Material Considerations**

   As with ion and electron accelerators, the beam energies involved are sufficient to generate ionizing radiation such as bremsstrahlung x-rays. To provide a safe environment for students and instructors, sufficient shielding may be needed depending on the operating voltage and beam current. For D-D operation at 30 keV and beam current of 0.05 mA, we measured the x-ray dose to be 0.2 μSv/hr at approximately 5cm from the target area. A more in-depth scan around the ion beam system shows that the x-ray dose is between 0.15 to 0.25 μSv/hr at 5 cm standoff distance. The background dose by comparison is ~0.07 μSv/hr. We note that the elevated x-ray dose is only present at the close standoff distance. Once the standoff distance is increased to ~20 cm, which is a typical minimum distance at which users will interact with the device, the x-ray dose level has reached the background level.

   When operating at 75 keV for the proton beam, we did observe a high count of bremsstrahlung x-ray. However, after attaching a double layered 0.5 mm thick flexible lead blanket around the ion beam system, the x-ray dose was effectively reduced to the background level. Furthermore, we did not detect radiation above the background level in the target area.

Thus, we can conclude that thin lead shielding will only be necessary around the acceleration region when operating at the higher beam energies.

In addition to the x-ray dose, we can also estimate the neutron radiation dose based on the charged particle count rate on the Si PIN detector. The total counts on the detector at 40 keV was ~44000 over 267 seconds, or ~165 counts per second. The detector area is 0.9 x 0.9 cm$^2$ at a distance of 4.445 cm thus covering a fraction of the total solid-angle equal to 0.00326. Since the detector counts all three charged particles (helion, triton, proton), only 1/3 of which will correlate with neutron production. Thus, the total neutron yield of the ion beam system can be calculated to be ~16900 n/s. The anticipated minimum user stand-off distance during operation to be >50cm from the neutron point source, which is a reasonable distance for the user to be at relative to the ion beam system. At this distance, the neutron flux approximates to ~0.54 n/cm$^2$/s. In order to estimate the neutron radiation dose to biological tissue, we use the dose conversion factor for neutrons at 2.5 MeV energy based on ICRP-74 [28] is ~400 pSv·cm$^2$. Multiplying the neutron flux by the dose conversion factor and changing units, one can obtain the neutron dose rate in μSv/hr, which approximates to ~0.77 μSv/hr (0.077 mrem/hr). According to the Nuclear Regulatory Commission (NRC), the recommended radiation dose to the public should not exceed 1 mSv per year [29]. Based on the level of neutron radiation produced by the ion beam system, a person would need to stay within 50 cm of the device for more than 1300 hours in a year to exceed the recommended dose. Given that the average work hours per year is ~2000 (40 hrs/week at 50 weeks), we do not think this dose will be exceeded for nominal operation of the device.

As with all neutron generating devices, one should always be cognizant of any radioactive material that is generated due to neutron activation. For the ion beam system, we have identified two materials of interest which are worthy of a more in-depth calculation to determine its activation hazard. The first is Cr-50 from the stainless steel 6-way cube surrounding the target and the second is Al-27 on the target disk. These two materials could potentially activate via. the (n,γ) process into Cr-51 and Al-28, which emit gamma rays of energies 0.32 MeV and 1.78 MeV respectively. For Cr-51, the half-life is 27.7 days and has a large thermal neutron activation cross-section, which could be considered a long-term radioactive hazard; whereas Al-28 should be treated as a short-term hazard given its 2.2 minute half-life.

For the activation calculation, we use the equation given in [31] to estimate the maximum possible gamma activity from the activated material:

$$A = N \cdot \sigma \cdot \Phi \cdot b \cdot \left(1 - e^{-\lambda \cdot t_a}\right) \cdot e^{-\lambda \cdot t_w} \qquad (6)$$

Here, N is the total number of target atoms in the volume of material, σ is the (n,g) activation cross-section at thermal neutron energies (~0.025 eV) taken from the ENDF/B-VIII.0 database [32], Φ is assumed to be the average thermal neutron flux at the center-of-mass of the volume of material, b is the branching ratio of gammas emitted per disintegration, λ is the half-life, $t_a$ is the irradiation time, and $t_w$ is the waiting time after irradiation. The number of target atoms can be calculated from the mass of the stainless-steel cube (1.25 kg) as:

$$N = f_{Cr} \cdot n_{Cr50} \cdot m \cdot \frac{N_A}{M_{Cr}} \qquad (7)$$

Where $f_{Cr}$ is the fraction of chromium in 304L stainless steel, $n_{Cr-50}$ is the natural abundance of Cr-50, m is the total mass, $N_A$ is Avogadro's number and $M_{Cr}$ is the molar mass of chromium. The steady-state activity can be achieved after irradiation for 5 half-lives with no wait time, resulting in approximately 99% of the maximum possible activity. Additionally, we assumed a thermal neutron to fast neutron fraction of 0.5, which is an overestimate considering the absence of neutron moderator material nearby. Nevertheless, this assumption provides an upper bound for the potential radiological material production. The maximum Cr-51 gamma activity is calculated to be ~8 Bq (0.2 nCi) and distributed over the entire volume of the stainless-steel cube. Compared to a typical 1 µCi Cs-137 check source emitting 0.662 MeV gammas, the amount of Cr-51 present in the ion beam system is more than 3 orders of magnitude lower, which should not be any cause for concern for regulatory compliance. Similarly, the amount of Al-28 generated in the target was estimated to be ~100 Bq (2.7 nCi) after 11 minutes of irradiation, which is 2 orders of magnitude lower than the Cs check source. After ~10 half-lives (~20 minutes), the amount of Al-28 would have decayed to background levels.

**Conclusion**

In summary, we have presented a compact ion beam system that is tailored to be used in an educational setting where students and users can get a first-hand experience studying the fundamental fusion reactions (e.g. D-D, p-B11). Ions are accelerated from a microwave driven plasma which bombards a selected target material to produce various fusion reactions. Students will be able to use a combination of solid-state electronic detectors and CR-39 plastic for charged particle detection. Simple cloud chambers can be attached to visualize the 3.02 MeV proton taking advantage of its ability to penetrate the thin vacuum membrane. In addition to studying nuclear reactions using the ion beam system, students are also exposed to the various technologies that drive the system (e.g. microwave engineering, high voltage insulation, etc.) This is a system which should find a lot of appeal in both undergraduate and graduate programs for STEM education.

**Acknowledgement**

The authors are grateful for discussions with Prof. Roger Falcone, Prof. Richard Petrasso, Prof. John Martinis, Prof. Jyhpyng Wang, Dr. Hao-Lin Chen, Dr. Kosta Yanev, Mr. Paul Chau, Ms. Fay Li, Mr. Peter Liu, Dr. Mason Guffey, Dr. David Chu, Mr. KaiJian Xiao, Dr. Peter Hsieh, Dr. Steve Hwang, Mr. Wilson Wu, Mr. Charles Wu, Mr. David Noriega, Mr. Ryan Yan, Ms. Lilly Zhang, Ms. Belinda Mei. This work was conducted by scientists and engineers as employees of Alpha Ring International Limited and its affiliated companies. The study was funded by Alpha Ring International Limited, which supports the education and training of a fusion industry workforce. The funding from Alpha Ring International Limited did not influence the scientific integrity or the results of the study.